\documentclass[prl,a4paper,twocolumn,showpacs,superscriptaddress,floatfix]{revtex4}
\usepackage{amsmath,amsthm}
\usepackage{amsfonts}
\usepackage[hmargin=2cm,vmargin=2cm,bmargin=2cm]{geometry}
\usepackage{graphicx}

\theoremstyle{definition}

\theoremstyle{remark}

\usepackage[latin1]{inputenc}
\usepackage{graphicx}
\usepackage{braket}
\usepackage{pdfpages}
\usepackage{tensor}
\usepackage{epstopdf}
\begin{document}
\title{Parity and Time Reversal Symmetry in Hanbury Brown-Twiss Effect}
\author{E. R. Silva}
\affiliation{Departamento de F\'isica, Universidade Federal da Para\'iba, 58051-970, Jo\~ao Pessoa, Para\'iba, Brazil.}
\author{A. L. R. Barbosa}
\affiliation{Departamento de F\'isica, Universidade Federal Rural de Pernambuco, Dois Irm\~aos, 52171-900 Recife, Pernambuco, Brazil}
\author{J. G. G. S. Ramos}
\affiliation{Departamento de F\'isica, Universidade Federal da Para\'iba, 58051-970, Jo\~ao Pessoa, Para\'iba, Brazil.}
\date{\today}
\begin{abstract}

The current manuscript employs the parity and time reversal symmetry in the Hanbury Brown-Twiss experiment. For this purpose, we develop a general scattering matrix framework founded on the concatenation of many individual compounded scattering processes on the setup. In this way, we derive the general scattering matrix of a parity and time reversal symmetric Hanbury Brown-Twiss experiment (HBT-PT). Within such scattering formulation, we propose a theoretical framework which provides how to measure the symmetry of the system through the correlation function of a pair of particles transmitted through the leads. The correlation function naturally reveal the quantum statistics of both bosons and fermions and demonstrate a very preponderant role of PT symmetry on the HBT experiment. We indicate the formation of both quantum and classical universal Turing machine depending on controllable parameters of the apparatus.

\end{abstract}
\maketitle

\section{I. Introduction}

Over the last few years, parity and time reversal (PT) symmetric Hamiltonians have been consolidated as a possible extension of the Hermitian Hamiltonian postulate of quantum mechanics \cite{ref1}, a Dirac proposal to obtain real energy eigenvalues \cite{ref2}. Systems described by a PT-symmetric Hamiltonian are tested since then in different scenarios. For instance, in condensed matter, a myriad of complex PT lattices generates real energy bands \cite{ref3}, and, in the Hatano-Nelson model, the typical non-Hermitian Hamiltonian has been used to analyze the flux lines of a magnetic field in a type-II superconductor \cite{ref4}. Also, in quantum optics, the systems can be PT-symmetrically adjusted in a balanced section of a threshold laser with a coherent perfect absorber, CPA \cite{ref5}. Henceforth, in order to study the eigenvalue problem of a non-Hermitian system, we should identify its PT transformation.

Exploring the non-Hermitian Hamiltonian postulate, Schomerus investigates interesting properties of a PT-invariant one-dimensional resonator \cite{ref8} and shows how to detect the PT-symmetry in such system using transport properties. The main underlying mechanism is the preservation of the number of particles in a quantum scattering through a simultaneous loss and gain system combined with PT symmetry. In the Hermitian configuration, the system conventionally supports quasi-bound states with the lifetime estimated by the imaginary term of the self-energy, but slip away to support stationary states for non-Hermitian Hamiltonian. Schomerus demonstrates the PT-symmetry fingerprints of the Hamiltonian through: the quantum noise, the stability at the lasing threshold, the emitted radiation, etc.

Motivated by the Schomerus ideas, we investigate how the shot-noise power, embedded in the amplitude of the temporal correlation function of multi-lead scattering, can reveal not only the nature of the particles, but also the PT-symmetry itself and its consequences. In the midst of all possible multi-lead experiments, we choose a cornerstone of optics, the Hanbury Brown-Twiss (HBT) interferometer.

The HBT experiment is due to R. Hanbury Brown and R. Q. Twiss \cite{ref11}. They propounded an interferometer that measures the intensity between particles \cite{ref6} emitted, with a time delay, by two photomultipliers. The incident particles impinge on the mirror adjusted in a way that the particles split in two common components with their own particle detectors. Thus, Hanbury Brown and Twiss experiment permits to find a bunching effect between photons as they reached the detectors \cite{ref12}, determining some fingerprints on the correlation between the delayed photons emitted in the apparatus. Similarly, the transversal electronic transport of particles in an edge state of the quantum Hall effect \cite{ref15} produces an anti-bunching effect \cite{ref13}. In the former case, the HBT experiment is founded on bosonic particles, and, in the latter, it is founded on fermionic ones. Then, we recognize the HBT experiment as a mechanism to measure the statistical behavior of correlated particles. This ubiquitous possibility to measure the statistics of fundamental particles encouraged R. Glauber to formulate a general quantum optics concept \cite{ref14} and other applications, $e.g.$ superconductivity (one can establish an isometry between a Cooper pair ray and the HBT experiment \cite{ref16}), elementary particles (pion decay resulting of nuclear collisions \cite{ref17}), solid state physics (waves suffering Anderson localization are correlated according HBT experiment \cite{ref18}) and quantum information (measuring the Bell's inequality violation \cite{ref25}).

Therefore, we explore in this paper the emerging physical mechanisms  attributed to the parity and time reversal symmetry transformation. With this motivation, we analytically derive the resulting scattering matrix due to both the coupling of a tunneling probability barrier with arbitrary transmission parameter and to an amplifying/absorber system in each HBT's lead. The compound system will allow ones to consider the phenomena as a result of resonant effects. As we derive the scattering matrix, denoted as HBT-PT, we analytically calculate the correlation function which relates the incident/scattered particles with the quantum statistics of the system. The final result allow ones to establish the physical requirements to detect the PT-symmetry on the HBT-PT system. Also, we indicate some remarkable physical consequences.

\section{II. Scattering Matrix}

In the current section, we deduce both the HBT (Hermitian) and the HBT-PT (non-Hermitian) scattering matrices. The first deduction is a revision to introduce the reader on the subject. The second deduction is an original result inspired on the Schomerus ideas and manifests some peculiarities.

\begin{figure}[h]
\centering
\includegraphics[scale=0.4]{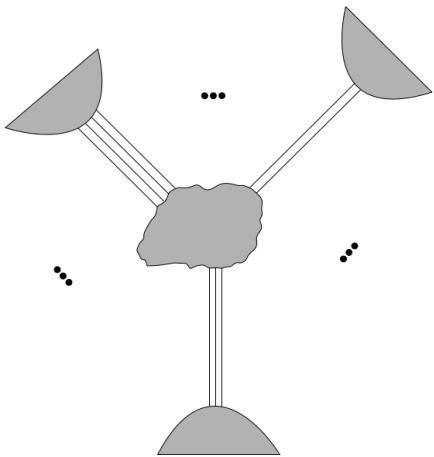}
\caption{Scattering center coupled to reservoirs through ideal leads.}
\label{figura1}
\end{figure}

\subsection{II. (a) Hermintian Hambury Brown-Twiss Setup}

For pedagogical reasons and in order to render the paper self-contained, we first explore the conventional HBT experiment, exhibiting how the amplitude of the temporal correlation function of counting statistics (shot-noise) is directly related to the nature of the particles. We consider now a generic system with $\alpha$ ideal leads coupled to a scattering center (SC) as depicted in Fig.(\ref{figura1}). The leads will transport particles (fermions or bosons) from the reservoirs to the SC, and also from the SC to the detectors. For this reason, the wave solution on the {\bf two}-dimensional lead $\alpha$ can be written as
\begin{equation}\label{eq1}
\psi_\alpha (x,y) = \displaystyle\sum_{n = 1}^{N_\alpha}(a_n^{(\alpha)} \psi_n^{- (\alpha)} + b_n^{(\alpha)} \psi_n^{+(\alpha)}),
\end{equation}
where the SC is the reference point. The coefficients $b_n^{(\alpha)}$ and $a_n^{(\alpha)}$ are the output and input amplitudes, respectively, while $N_\alpha$ is the number of open channels in the $\alpha$-th lead. The functions $\psi^{\pm(\alpha)}_n$ are plane waves as we assume very far detectors (and emitters) from the SC. We can represent the scattering formulation of Eq.(\ref{eq1}) assuming that the output amplitudes are the result of an unitary transformation of the input amplitudes,
\begin{equation}\label{eq2}
B=SA,
\end{equation}
where $A$ and $B$ stands for the column vector of input and output amplitudes, respectively, and $S$ is the scattering matrix which provides the relation between these amplitudes. The $S$ matrix is unitary, i.e., $SS^{\dagger}=\mathbb{I}$ in order to preserve the probability current. The scattering matrix fulfills all constraints because it accounts all possible interactions that affect the transport properties of the system. The entries of the scattering matrix are usually written as

\begin{equation} \label{ScM}
S =
\left( \begin {array}{cc} \mathbf{r}&\mathbf{t}^\prime\\ \noalign{\medskip}\mathbf{t}&\mathbf{r}^\prime\end {array}
 \right),
\end{equation}
with $\mathbf{r}$, $\mathbf{r}^\prime$ and $\mathbf{t}$, $\mathbf{t}^\prime$ representing the reflection and transmission blocks of the scattering, respectively.

\begin{figure}[!t]
\centering
\includegraphics[scale=0.4]{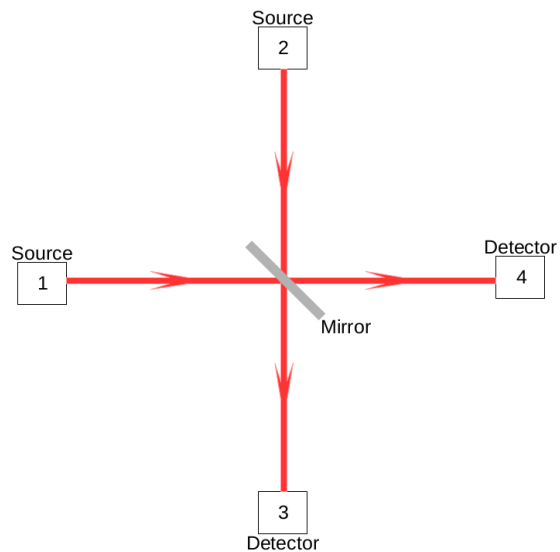}
\caption{Hanbury Brown-Twiss interferometer.}
\label{figura2}
\end{figure}

The scattering formalism can be applied to the simplest HBT setup, consisting of two incident particles in different leads, each one with a single open channel (energy). As previously mentioned, Hanbury Brown and Twiss proposed an interferometer that allows to measure the correlation between two particles emitted by different leads into a semi-transparent mirror. The latter can be interpreted as a scattering center which allows the particles to be transmitted or reflected to detector leads, as depicted in the Fig.(\ref{figura2}).
The mirror does not allow the reflection to the incident channel. Following the Fig.(\ref{figura2}), the direct (perpendicular) switch from the source to the detector will be called henceforth as transmission (reflection), with amplitude denoted as $t$ ($r$). Accordingly, the conditions that reproduce the HBT setup through the Eq.(\ref{eq2}) is
\begin{eqnarray} \label{eq3}
&\left( \begin{array}{c}
b_1\\
b_2\\
b_3\\
b_4
\end{array} \right)=\left( \begin{array}{cccc}
0 & 0 & r^* & t^* \\
0 & 0 & t^* & r^* \\
r& t & 0 &0\\
t & r &0 &0
\end{array} \right)\left( \begin{array}{c}
a_1\\
a_2\\
a_3\\
a_4
\end{array} \right),&\nonumber\\
&\left( \begin{array}{c}
b_1\\
b_2\\
b_3\\
b_4
\end{array} \right)=S_\textrm{HBT}\left( \begin{array}{c}
a_1\\
a_2\\
a_3\\
a_4
\end{array} \right),&
\end{eqnarray}
with $r$ and $t$ expressing the reflection and transmission amplitudes of the mirror, respectively. The $4 \times 4$ scattering matrix in Eq.(\ref{eq3}) has the backscattering properties, allowing the permutation $source \leftrightarrow detector$. This permutation is another interesting fact presented in $S_\textrm{HBT}$ and reveals that it is Hermitian. For our conceptual proposal, without loss of generality, we set the mirror to reflect and transmit with the same probabilities amplitudes, i.e., $-r=it=\sqrt{2}/2$. Therefore, the second quantization techniques \cite{ref7} allow to find the detection probabilities of the two particles in the same lead, $P(ii), \; i=3,4$, or in different leads, $P(i,j), \; i \neq j =3,4$. The result is
\begin{eqnarray}
&\textrm{P}(33)=\textrm{P}(44)= \dfrac{1}{4}(1-\epsilon |I|^2)& \label{eq4},\\
&\textrm{P}(34)= \dfrac{1}{2}(1+\epsilon |I|^2)\label{eq5}.&
\end{eqnarray}

The Eqs.(\ref{eq4},\ref{eq5}) are valid for both fermions and bosons, while the parameter $\epsilon$ informs the two possibles algebras used: if $\epsilon=1$, the algebra is fermionic, and, if $\epsilon=-1$, the algebra is bosonic. Furthermore, $I$ is the overlap between the incident particles waves which mutually correlates due to their indistinguishability. Experimentally, the overlap parameter $I \in [0,1]$ encodes the simultaneity of the emitted particles: if $I=1$, the particles are emitted exactly at the same time, and, if $I=0$, the particles are emitted with a sufficiently large time delay. Despite the present system is the simplest example of a HBT experiment, notice it manifests the Pauli exclusion principle for the perfect overlapped fermionic waves as the probability to find the two particles in the same lead is zero from Eq.(\ref{eq4}). The opposite behavior occurs for the bosonic waves according to the same equation, resulting in the bunching effect.

\subsection{II. (b) Non-Hermintian Hambury Brown-Twiss Setup}

\begin{figure}[!t]
\centering
\includegraphics[scale=0.13]{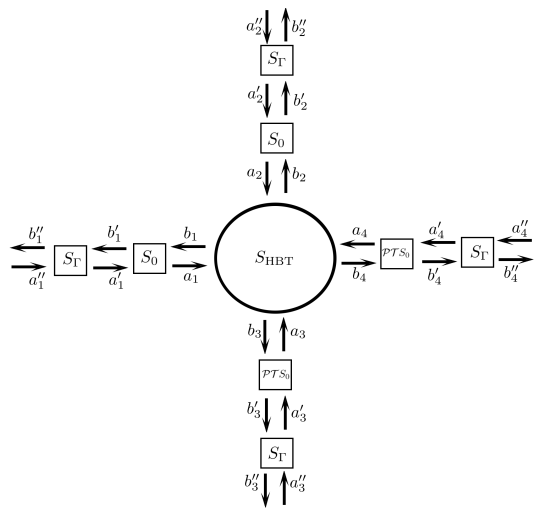}
\caption{Amplifying-Absorber sections and barriers coupling to the HBT apparatus\cite{ref8}.}
\label{figura3}
\end{figure}

Concluded a brief presentation of the conventional Hermitian HBT setup, we begin the HBT-PT formal analysis. The PT-symmetry in the HBT experiment can be controlled through a amplifying-absorber mechanism \cite{ref8} which modulates the parity and time reversal symmetry breaking. Therefore, performing an extension of the one-dimensional setup \cite{ref8}, we couple in usual way two orthogonal amplifying sections in the leads $1$ and $2$, generating the following scattering matrices (Fig.\ref{figura3})
\begin{eqnarray} \label{eq6}
&\left( \begin{array}{c}
b^\prime_i\\
a_i
\end{array} \right)=\left( \begin{array}{cc}
0 & t_0 \\
t_0 & 0
\end{array} \right)\left( \begin{array}{c}
a^\prime_i\\
b_i
\end{array} \right),&\nonumber \\
&\left( \begin{array}{c}
b^\prime_i\\
a_i
\end{array} \right)=S_0\left( \begin{array}{c}
a^\prime_i\\
b_i
\end{array} \right),&
\end{eqnarray}
where $t_0$ is a experimental complex parameter which rules the amplifying rate section and $i=1,2$. The ${\cal T}$-symmetry can be broken basically through two mechanisms: a perpendicular magnetic field or some amplification/absorption parameter, both contributing with a imaginary term to the Hamiltonian $H$. {\bf Such Hamiltonian become non-hermitian when we consider open systems, as the case of amplification/absorption sections. It allows one to talk about non-Hermicity of well-established Hermitian systems, as two-dimensional electron transport\cite{ref27}}. The effect of the ${\cal T}$ broken is algebraically identified with the operation $\mathcal{T}S_{0}=(S_0^{*})^{-1}$. On the other hand, a parity ${\cal P}$ transformation modifies the wave function $\psi(x)$ as $\mathcal{P} \psi (x)=\psi(-x)$, which exchanges the left/right or up/down leads as $\mathcal{P}S_{0}=\sigma_x S_{0} \sigma_x$. Therefore, the PT-symmetry is preserved if the diametrically opposite (parity) lead balance the two absorber sections in the leads $3$ and $4$. The resulting scattering equation is the result of the parity and time reversal operators, $\mathcal{P}$ and $\mathcal{T}$, respectively. We implement the procedure on $S_0$, following the Ref.\cite{ref8} and inspecting the setup depicted in the Fig.(\ref{figura3}) as
\begin{eqnarray} \label{eq7}
&\left( \begin{array}{c}
b^\prime_{i+2}\\
a_{i+2}
\end{array} \right)=\mathcal{PT}S_0\left( \begin{array}{c}
a^\prime_{i+2}\\
b_{i+2}
\end{array} \right),&\nonumber \\
&\left( \begin{array}{c}
b^\prime_{i+2}\\
a_{i+2}
\end{array} \right)=\sigma_x (S^*_0)^{-1} \sigma_x\left( \begin{array}{c}
a^\prime_{i+2}\\
b_{i+2}
\end{array} \right),&\nonumber \\
&\left( \begin{array}{c}
b^\prime_{i+2}\\
a_{i+2}
\end{array} \right)=\left( \begin{array}{cc}
0 & \frac{1}{t_0^*} \\
\frac{1}{t_0^*} & 0
\end{array} \right)\left( \begin{array}{c}
a^\prime_{i+2}\\
b_{i+2}
\end{array} \right),&
\end{eqnarray}
where $\sigma_x$ is the Pauli matrix. The Eq.(\ref{eq7}) is the expected resonant effect result \cite{ref19}. The complete coupling of the HBT-PT setup sections is achieved with the substitution of $a$ and $b$ of Eq.(\ref{eq6}) and Eq.(\ref{eq7}) in the Hermitian HBT scattering equation given in the Eq.(\ref{eq3}). The final result is
\begin{eqnarray}\label{S+Sc}
&\left( \begin{array}{c}
b^\prime_1\\
b^\prime_2\\
b^\prime_3\\
b^\prime_4
\end{array} \right)=-\dfrac{\sqrt{2}}{2}\dfrac{t_0}{t_0^*}\left( \begin{array}{cccc}
0 & 0 & 1&-i \\
0 & 0 &-i& 1 \\
 1&  i& 0& 0\\
 i&  1& 0& 0
\end{array} \right)\left( \begin{array}{c}
a^\prime_1\\
a^\prime_2\\
a^\prime_3\\
a^\prime_4
\end{array} \right).&
\end{eqnarray}
Notice the scattering matrix in Eq.(\ref{S+Sc}) is no longer Hermitian as in Eq.(\ref{eq3}), although Eq.(\ref{S+Sc}) still reproduce the same outcome of the standard (Hermitian) HBT setup, as we show in the section IV. The Hermitian case is recovered if $\textrm{Im}(t_0)=0$. Therefore, the imaginary part of $t_0$ is responsible for the crossover between the experiments HBT and the HBT-PT. {\bf Another subtle point is about the noise created by inserting the amplifying/absorbing sections, once this clearly changes the number of photons detected or a attenuation in the signal of an electron transported, for example. We can work around this applying tunneling barriers in each lead and explore only the resonant regime of the system, then we guarantee to detect only transmitted particles with the specifics quantization conditions of the system \cite{ref28}.} As depicted in the Fig.(\ref{figura3}), we couple a tunneling barrier in the leads in order to explore the resonant regime. The scattering produced by the barriers can be written as
\begin{eqnarray} \label{eq8}
&\left( \begin{array}{c}
b^{\prime\prime}_j\\
a^\prime_j
\end{array} \right)=-\left( \begin{array}{cc}
\sqrt{1-\Gamma} & i\sqrt{\Gamma} \\
i\sqrt{\Gamma} & \sqrt{1-\Gamma}
\end{array} \right)\left( \begin{array}{c}
a^{\prime\prime}_j\\
b^\prime_j
\end{array} \right),&\nonumber \\
&\left( \begin{array}{c}
b^{\prime\prime}_j\\
a^\prime_j
\end{array} \right)=S_\Gamma\left( \begin{array}{c}
a^{\prime\prime}_j\\
b^\prime_j
\end{array} \right),&
\end{eqnarray}
where $\Gamma \in [0,1]$ is the transmission probability generated by the barriers and $j=1,2,3,4$. Proceeding in the usual way, we substitute $a^\prime$ and $b^\prime$ in Eq.(\ref{S+Sc}) and, after some algebra, we find a considerably simplified form to the total scattering matrix
\begin{equation}\label{eq9}
S_\textrm{T} =  -\dfrac{1}{(1-\Gamma)\left(\frac{t_0}{t_0^{*}}\right)^2-1}\left( \begin{array}{cccc}
s & 0 & s^\prime & -is^\prime \\
0 & s & -is^\prime & s^\prime \\
s^\prime & is^\prime & s & 0 \\
is^\prime & s^\prime & 0 & s
\end{array} \right),
\end{equation}
where
\begin{eqnarray}
&s \equiv \sqrt{1-\Gamma}\left[\left(\frac{t_0}{t_0^{*}}\right)^2-1\right] ,& \label{eq10}\\
&s^\prime \equiv  -\dfrac{\sqrt{2}}{2}\dfrac{t_0}{t_0^*}\Gamma.\label{eq11}&
\end{eqnarray}
The Eq.(\ref{eq9}) represents the scattering matrix which combines the HBT experiment with the control mechanisms of parity and time reversal symmetries (HBT-PT). One may argue that the barriers would affect the symmetry in each lead of the Eq.(\ref{S+Sc}), but it is not true since $S_\Gamma$ is invariant under $\mathcal{PT}$ transformation, $S_\Gamma=\mathcal{PT}S_\Gamma$, as one can verify. The scattering matrix $S_\textrm{HBT}$ can be obtained from the Eq.(\ref{eq9}) if Im$(t_0)=0$ and $\Gamma\rightarrow 1$, as expected. The resonant regime is achieved if Im$(t_0)=0$ and $\Gamma\rightarrow 0$, generating singularities in the output amplitudes with the consequent parity and time reversal symmetry breaking.

Although the representation of the system through the scattering matrix of the Eq.(\ref{eq9}), a experimental setup to measure the quantum statistics is highly non-trivial if the apparatus is symmetrically adjusted to satisfy parity and time reversal. A method would be the system immersion in a heat bath \cite{ref8, ref20, ref21}, from which we evaluate the system response using the noise due to the bath-system coupling and the consequent fluctuation-dissipation \cite{ref22}. This provides a analysis of the attenuation of the output amplitudes \cite{ref23}. Clearly this method is very hard to implement in the four leads detection apparatus. Fortuitously, we formulated an alternative approach inspired on the Schomerus ideas to determine the PT-Symmetry using counting statistics, a original general formalism develop in next section for multi-lead systems.

\section{III. Input and output states.}

In order to provide the formalism, it is convenient to write the inputs in terms of the outputs
\begin{equation}\label{eq12}
A=S^\dagger B.
\end{equation}
We consider a Multi Singled Channel Lead System (MSCTS) operating at zero temperature. The scattering matrix of such system can be written as
\begin{equation}\label{eq13}
S_\textrm{MSMTS}= \left( \begin {array}{cccc} r_{{11}}&t_{{12}}&{\cdots }&t_{1\alpha}
\\ \noalign{\medskip}t_{{21}}&r_{{22}}&{\cdots }&t_{2\alpha}
\\ \noalign{\medskip}{\vdots }&{\vdots }&{\ddots }&{\vdots }
\\ \noalign{\medskip}t_{\alpha1}&t_{\alpha2}&{\ldots }&r_{\alpha\alpha}
\end {array} \right).
\end{equation}
As each lead has only one open channel, it is simple to express a field operator that creates a particle in the lead $\alpha$ with wave function $\psi_i(\vec{\textrm{r}}_\alpha)$ \cite{ref9},
\begin{equation} \label{eq14}
\hat{\psi}^{\dagger}_{\alpha i} = \int \textrm{d}\vec{\textrm{r}}_{\alpha} \psi_i(\vec{\textrm{r}}_{\alpha}) \hat{a}^{\dagger}(\vec{\textrm{r}}_{\alpha}).
\end{equation}
We denote $N$ as the number of leads in the system while $\hat{a}^{\dagger}(\vec{\textrm{r}}_{\alpha})$ is the standard creation operator. The field operator of the Eq.(\ref{eq14}) will act in multiplet states of the Fock space. On the other hand, the field operator will obey the following commutation relations
\begin{eqnarray}
&\left[\hat{\psi}_{\alpha i}, \hat{\psi}^{\dagger}_{\alpha i} \right]_\epsilon = 1,&\label{eq15} \\
&\left[\hat{\psi}_{\alpha i}, \hat{\psi}^{\dagger}_{\alpha j} \right]_\epsilon =\left[\hat{\psi}_{\alpha j}, \hat{\psi}^{\dagger}_{\alpha i} \right]^*_\epsilon =  I,&\label{eq16} \\
& \left[\hat{\psi}_{\alpha i}, \hat{\psi}_{\beta j} \right]_\epsilon =\left[\hat{\psi}_{\alpha i}, \hat{\psi}_{\beta i} \right]_\epsilon = 0,&\label{eq17}
\end{eqnarray}
where $I$ is the same overlap factor of the Eqs.(\ref{eq4}) and (\ref{eq5}) that are defined in terms of the wave functions as
\begin{equation} \label{eq18}
I = \int \textrm{d}\vec{\textrm{r}}_{\alpha} \psi_i^*(\vec{\textrm{r}}_{\alpha}) \psi_j(\vec{\textrm{r}}_{\alpha}).
\end{equation}

The possible $2$ incident particle states can be constructed using the field operator of the Eq.(\ref{eq14}) and normalizing the fields. After some algebra, we obtain the states
\begin{eqnarray}
&\left| \alpha \alpha \right>  = \left[1 - \epsilon |I|^2\right]^{-1/2} \hat{\psi}^{\dagger}_{\alpha i} \hat{\psi}^{\dagger}_{\alpha j} \left|0 \right>,&\label{eq19} \\
&\left| \alpha \beta \right>_\pm = \left[2(1 \pm  |I|^2)\right]^{-\frac{1}{2}}\left(\hat{\psi}^{\dagger}_{\alpha i} \hat{\psi}^{\dagger}_{\beta j} \pm \hat{\psi}^{\dagger}_{\alpha j} \hat{\psi}^{\dagger}_{\beta i} \right) \left|0 \right>.\label{eq20}&
\end{eqnarray}
The relations expressed in the Eqs.(\ref{eq19}) and (\ref{eq20}) form the input states. They can be referred as the singlet and doublet states, respectively. The latter contemplates the symmetric and antisymmetric ones. The particles interact in the SC and will be created in any other lead preserving the constraints imposed by the conservative laws on the scattering matrix. The relation between the input field operators and the output ones follow from the Eq.(\ref{eq12})
\begin{equation}\label{eq21}
\hat{\psi}^{\dagger}_{\alpha i} = r_{\alpha \alpha} \hat{\psi}^{\dagger}_{\alpha i} + \displaystyle\sum_{a}^N t_{\alpha a}\hat{\psi}^{\dagger}_{ai},
\end{equation}
We substitute the Eq.(\ref{eq21}) in the Eqs.(\ref{eq19}) and (\ref{eq20}) concomitantly with the commutation relations (\ref{eq15})-(\ref{eq17}) and obtain the novel identity
 \begin{widetext}
\begin{eqnarray}\label{eq22}
&\displaystyle\sum_{a,b}t_{\alpha a}t_{\beta b}\hat{\psi}^{\dagger}_{a i}\hat{\psi}^{\dagger}_{b j} = \displaystyle\sum_{a}\left[t_{\alpha a}t_{\beta a}\hat{\psi}^{\dagger}_{a i}\hat{\psi}^{\dagger}_{a j} + \displaystyle\sum_{n}\left( t_{\alpha a}t_{\beta a+n}\hat{\psi}^{\dagger}_{a i}\hat{\psi}^{\dagger}_{a+n j} + t_{\alpha a+n}t_{\beta a}\hat{\psi}^{\dagger}_{a+n i}\hat{\psi}^{\dagger}_{a j} \right) \right].&
\end{eqnarray}
\end{widetext}
We rearrange the terms of the resultant expression to find all the output Fock states. After some algebra, the input states in terms of output states given through the novel general MSCTS equations
  \begin{widetext}
\begin{eqnarray}
\left|\alpha \alpha \right> &=& r^2_{\alpha\alpha}\left| \alpha \alpha \right> + \sqrt{2}r_{\alpha\alpha} t_{\alpha a} \left| \alpha a \right>_{-\epsilon}  + t^2_{\alpha a} \left| aa \right> + \sqrt{2} t_{\alpha a}t_{\alpha a+n} \left| aa+n\right>_{-\epsilon} , \label{eq23} \\
\left| \alpha\beta \right>_\pm&=& \mathcal{I}^\pm_{\epsilon}\sqrt{(1\mp \epsilon)}\left( r_{\alpha\alpha} t_{\beta \alpha} \left| \alpha\alpha \right>+r_{\beta\beta}t_{\alpha\beta}\left|\beta\beta \right>+t_{\alpha a}t_{\beta a} \left|a a \right> \right)+\nonumber \\
&&+\left( r_{\alpha\alpha}r_{\beta\beta} \mp \epsilon t_{\alpha\beta}t_{\beta\alpha} \right)\left| \alpha \beta \right>_\pm+\left(r_{\alpha\alpha}t_{\beta a} \mp \epsilon t_{\alpha a}t_{\beta\alpha}\right)\left|\alpha a\right>_\pm +  \nonumber \\
&&\left( r_{\beta\beta}t_{\alpha a}\mp \epsilon t_{\alpha \beta}t_{\beta a} \right)\left|\beta a \right>_\pm+\left( t_{\alpha a}t_{\beta a+n} \mp \epsilon t_{\alpha a+n}t_{\beta a} \right)\left| a a+n\right>_\pm, \label{eq24}
\end{eqnarray}
\end{widetext}
where we define
\begin{equation}\label{eq25}
\mathcal{I}^\pm_{\epsilon} = \sqrt{\dfrac{1-\epsilon|I|^2}{(1\pm |I|^2)}}.
\end{equation}

It is necessary to fix some conventions implicit on the Eqs.(\ref{eq23}) and (\ref{eq24}). Firstly, we have omitted all the sums in the indices $a$ and $n$. To avoid misunderstandings, it is enough to consider that the index $a$ varies from $1$ to $N$ and $n$ varies from $1$ to $N-a$. Secondly, an important feature is that $\alpha$ and $\beta$ do not explicitly appear in the sum of $a$.  Given these conventions, we realize that the doublet input states is expressed by all the possible doublet output states. The formalism can be applied in a myriad of scenarios involving multi-lead scattering processes and, in particular, it will allow us to find the correlation functions of the HBT-PT system.

\section{IV. Correlation Function}

Given a doublet input state in a general system such as the one depicted on the Fig.(\ref{figura1}), we can determine the probability of the input state projection on each possible output state of the system. Furthermore, such probability can be interpreted as a correlation function provided the states are doublets. Accordingly, the counting statistics can be expressed in terms of correlations (noise) as
\begin{eqnarray}\label{eq26}
&|\left< \alpha \alpha|\textrm{Input} \right>|^2 = (\Delta n_\alpha)^2,& \\
\label{eq27}&|_{+}{\left< \alpha \beta|\textrm{Input} \right>}|^2+ |_{-}{\left< \alpha \beta|\textrm{Input} \right>}|^2= \left< n_\alpha n_\beta \right>, &
\end{eqnarray}
where $(\Delta n_\alpha)^2=P(\alpha \alpha)$ and $\left< n_\alpha n_\beta \right>=P(\alpha \beta)$ are the previously mentioned probabilities. The derivation of each possible relation between the input and output doublets expressed in Eqs.(\ref{eq23}) and (\ref{eq24}) is useful to construct some input state through $S^{\prime}$ and, consequently, to find the correlation functions through (\ref{eq26}) and (\ref{eq27}).

The Eqs.(\ref{eq19}) and (\ref{eq20}) can be cast into the form 
\begin{eqnarray}\label{eq28}
&\hat{\psi}^{\dagger}_{\alpha i} \hat{\psi}^{\dagger}_{\beta j} \left|0 \right> = \frac{1}{2}\left[2(1 - |I|^2)\right]^{1/2}\left| \alpha\beta \right>_- +&\\ \nonumber
&+ \frac{1}{2}\left[2(1 + |I|^2)\right]^{1/2}\left| \alpha\beta\right>_+ .&
\end{eqnarray}
With this set of equations, finally the correlation expressed on Eq.(\ref{eq26}) will be calculated. We substitute Eq.(\ref{eq24}) in the r.h.s of Eq.(\ref{eq28}) and multiply by one of the possible bras $\left< \alpha \alpha \right|$, $\left< \beta \beta \right|$ or $\left< aa \right|$. After some algebra and taking the square modulus, we get the correlation functions for an output lead of the same index
\begin{eqnarray}\label{eq29}
&(\Delta n_\alpha )^2 = |r_{\alpha\alpha}|^2|t_{\beta\alpha}|^2(1 - \epsilon |I|^2), & \\
\label{eq30}&(\Delta n_\beta )^2 = |r_{\beta\beta}|^2|t_{\alpha\beta}|^2(1 -\epsilon  |I|^2), & \\
\label{eq31}&(\Delta n_a )^2 = |t_{\alpha a}|^2|t_{\beta a}|^2(1 -\epsilon  |I|^2). &	
\end{eqnarray}

Proceeding similarly to $_{\pm}\left< \alpha \beta \right|$, $_{\pm}\left<\alpha a \right|$, $_{\pm} \left< \beta a \right|$ and $_{\pm} \left< a a+n \right|$, we can evaluate the Eq.(\ref{eq27}). The result is
\begin{widetext}
\begin{eqnarray}
\label{eq32} \left< n_\alpha n_\beta \right>&=&  |r_{\alpha\alpha}|^2|r_{\beta\beta}|^2 +  |t_{\alpha\beta}|^2|t_{\beta\alpha}|^2 -\epsilon |I|^2\left( r^*_{\alpha\alpha}r^*_{\beta\beta} t_{\alpha\beta}t_{\beta\alpha}+r_{\alpha\alpha}r_{\beta\beta} t^*_{\alpha\beta}t^*_{\beta\alpha}\right) . \\
\label{eq33}\left< n_\alpha n_a \right> &=& |r_{\alpha\alpha}|^2|t_{\beta a}|^2+ |t_{\alpha a}|^2|t_{\beta\alpha}|^2 - \epsilon |I|^2(r_{\alpha\alpha}t_{\beta a}t^*_{\alpha a}t^*_{\beta\alpha}+ r^*_{\alpha\alpha}t^*_{\beta a}t_{\alpha a}t_{\beta\alpha}), \\
\label{eq34}\left< n_\beta n_a \right> &=& |r_{\beta\beta}|^2|t_{\alpha a}|^2+ |t_{\beta a}|^2|t_{\alpha\beta}|^2 - \epsilon |I|^2(r_{\beta\beta}t_{\alpha a}t^*_{\beta a}t	^*_{\alpha\beta}+ r^*_{\beta\beta}t^*_{\alpha a}t_{\beta a}t_{\alpha\beta}), \\
\label{eq35}\left< n_a n_{a+n} \right> &=& |t_{\alpha a}|^2|t_{\beta a+n}|^2+ |t_{\alpha a+n}|^2|t_{\beta a}|^2 -\epsilon |I|^2(t_{\alpha a}t_{\beta a+n}t^*_{\alpha a+n}t^*_{\beta a}+ t^*_{\alpha a}t^*_{\beta a+n}t_{\alpha a+n}t_{\beta a}).
\end{eqnarray}
\end{widetext}
The Eqs.(\ref{eq29})-(\ref{eq35}) determine all the possible outcome of a generic MSCTS provided the corresponding scattering matrix. To our knowledge, this result, valid for any quantum system {\bf described by eqs. (\ref{eq1}) and (\ref{eq2}), is} completely novel. These equations immediately inform the shot-noise power of the system, the single source of noise at zero temperature. We extend our considerations to both fermionic and bosonic particles using the index $\epsilon$.

Now we focus on the HBT-PT scattering matrix, represented in the Eq.(\ref{eq9}). More specifically, the input indices on the defined HBT-PT are $\alpha=1$ and $\beta=2$. We substitute the scattering matrix elements in the correlation functions given above and obtain, after some algebra, the HBT-PT statistics
\begin{widetext}
\begin{eqnarray}
(\Delta n_1)^2&=&(\Delta n_2)^2=0, \label{eq36}\\
(\Delta n_3)^2&=&(\Delta n_4)^2= \dfrac{1}{4}\dfrac{\Gamma^4}{\left|(1-\Gamma)\left(\frac{t_0}{t^*_0} \right)^2 -1\right|^4}(1-\epsilon |I|^2), \label{eq37}\\
\left< n_1n_2 \right>&=& (1-\Gamma)^2\dfrac{\left|\left(\frac{t_0}{t^*_0} \right)^2 -1 \right|^4}{\left|(1-\Gamma)\left(\frac{t_0}{t^*_0} \right)^2 -1 \right|^4},\label{eq38}\\
\left< n_1n_3 \right>&=& \left< n_2n_4 \right>=\left< n_1n_4 \right>= \left< n_2n_3 \right>=\dfrac{1}{2}\Gamma^2(1-\Gamma)\dfrac{\left|\left(\frac{t_0}{t^*_0} \right)^2 -1 \right|^2}{\left|(1-\Gamma)\left(\frac{t_0}{t^*_0} \right)^2 -1 \right|^4},\label{eq39}\\
\left< n_3 n_4 \right>&=&\dfrac{1}{2}\dfrac{\Gamma^4}{\left|(1-\Gamma)\left(\frac{t_0}{t^*_0} \right)^2 -1 \right|^4}(1+\epsilon |I|^2).\label{eq40}
\end{eqnarray}
\end{widetext}

There is an interesting phenomenology in the Eqs.(\ref{eq36})-(\ref{eq40}). Firstly, we must note at this stage that in the absence of PT-symmetry, Im$(t_0)=0$, the standard (Hermitian) HBT effect result given in the Eqs.(\ref{eq4})-(\ref{eq5}) is recovered, independently of $\Gamma$. Therefore, the Hermitian HBT phenomenology is independent of resonant effects. On the other hand, in striking manner, the shot-noise power of the (non-Hermitian) HBT-PT apparatus strongly depends of the tunnel barriers $\Gamma$. Moreover, a remarkable fingerprint of the the non-Hermitian experiment is its non-null correlation functions between input and output leads as indicated on the Eq.(\ref{eq39}), despite the overlap term, $I$, occurs only on output arms in both cases, as indicated in the Eqs.(\ref{eq37}) and (\ref{eq40}). Secondly, and even more remarkable, the limit $\Gamma\rightarrow 1$ of the five last equations is
\begin{eqnarray}
&(\Delta n_1)^2=(\Delta n_2)^2=\left< n_{1,2}n_{2,3,4} \right>=0,&\label{eq41} \\
&(\Delta n_3)^2=(\Delta n_4)^2= \dfrac{1}{4}(1-\epsilon |\textrm{I}|^2),& \label{eq42}\\
&\left< n_3 n_4 \right>=\dfrac{1}{2}\left(1 +\epsilon |\textrm{I}|^2\right),\label{eq43}&
\end{eqnarray}
{\it i.e.}, the same result of the Eqs.(\ref{eq4}) and (\ref{eq5}). Therefore, the shot-noise power of both the Hermitian and non-Hermitian HBT is the same in the ideal system which once again indicates the preponderant role of tunneling in PT symmetric systems.

Given the dependence of the HBT-PT on the tunneling, we analyze the opaque limit, $i.e.$, impenetrable barriers ($\Gamma \rightarrow 0$), and we obtain
\begin{equation}
\left< n_1 n_2 \right>=1. \label{eq44}
\end{equation}
The Eq.(\ref{eq44}) states that the only possible output is the input doublet which projects in itself, indicating the full reflection of the incident particles. Finally, we explore details of the non-Hermitian HBT-PT correlations generated through amplifying and absorber sections. The results on the opaque regime can be summarized in simple expressions for both systems
\begin{eqnarray}
&(\Delta n_3)^2=(\Delta n_4)^2=\left\{ \begin{array}{l}\dfrac{1- \epsilon|\textrm{I}|^2}{4}, \textrm{if Im$(t_0)=0$} \\ 0, \textrm{ if Im$(t_0)\neq 0$}\end{array}\right. ,& \label{Turing1} \\
&\left< n_3 n_4 \right>=\left\{ \begin{array}{l}\dfrac{1+ \epsilon|\textrm{I}|^2}{2}, \textrm{ if Im$(t_0)=0$} \\ 0, \textrm{ if Im$(t_0)\neq 0$}\end{array}\right. ,& \label{Turing2} \\
&\left< n_1 n_2 \right>=\left\{ \begin{array}{l}0, \textrm{ if Im$(t_0)=0$} \\ 1, \textrm{ if Im$(t_0)\neq 0$}\end{array}\right. & \label{Turing3}
\end{eqnarray}
The exclusive sensitivity of non-Hermitian HBT-PT apparatus on the resonance phenomenona allow us to conclude that the shot-noise power in its typical regime is a very appropriate {\bf observable} of the PT-Symmetry. Also, there are possible technological applications if we consider a ``PT" gate similar to a transistor with alternating (input or output) $0$ shot-noise depending on the PT-symmetry of the system, as the previous equations indicate.

\section{V. Conclusion}

In the present paper, we investigate the amplitude of the temporal correlation function (shot-noise power) of both a Hanbury Brown-Twiss Hermitian and a non-Hermitian (PT) ones experiment.  The study is valid for bunching and anti-bunching effects of bosons and fermions, respectively. Inspired on the recent result of Schomerus, we consider the couple of amplifying-absorbing sections and also transmission barriers in the leads of the HBT apparatus. Through concatenation techniques, we obtain the resonant $S$-matrix of the HBT with parity and time reversal symmetry (non-Hermitian).

Henceforth, we extended the correlation functions formalism, by means of field operators defined in Fock space, which provide any statistics of a doublet input state of a multi-lead system, described by a generic scattering matrix. Once we derived the scattering matrix of the HBT-PT experiment, we obtain its statistics applying in the correlation functions. {\bf This latter} result can be applied in particular to arbitrary transmission probability, including the full opaque limit ($\Gamma \rightarrow 0$), and to arbitrary amplifying-absorbing rates, performing the full crossover to Hermitian and non-Hermitian (PT) statistics. Then we identify the breaking point of parity and time reversal symmetries, the fingerprint presented on the shot-noise power that separate completely the two regimes and can be used to the fine detection of the controlled PT symmetry and possibly to technological applications.

We hope that the developed multi-lead correlation function formalism permits to study another applications beyond the PT-Symmetric Hanbury Brown-Twiss effect.


\begin{thebibliography}{99}
\bibitem{ref6} Although in the original experiment was accomplished using fotons, later, it was verified the opposite effect for fermions, in this manner, we will refer to the carriers as particles.
\end{thebibliography}
\end{document}